\documentclass[intlimits,twoside,a4paper]{article}

\usepackage{amsmath,amssymb}
\usepackage{graphicx}
\usepackage{wrapfig}

\usepackage[T2A]{fontenc}
\usepackage[cp1251]{inputenc}

\usepackage[eqsecnum]{cmpj2}

\issue{2013}{16}{1}{13603}

\doinumber{10.5488/CMP.16.13603}

\title[Single-particle and pair condensates in Bose systems]%
{Role of single-particle and pair condensates in Bose systems with arbitrary intensity of interaction}%

\author[A.S. Peletminskii, S.V. Peletminskii, Yu.M. Poluektov]
{A.S. Peletminskii, S.V. Peletminskii, Yu.M. Poluektov }
\address{
Akhiezer Institute for Theoretical Physics, National Science Center
Kharkiv Institute of Physics and Technology,  1~Akademichna St., 61108~Kharkiv, Ukraine
}

\date{Received  April 27, 2012, in final form August 13, 2012}
\authorcopyright{A.S. Peletminskii, S.V. Peletminskii, Yu.M. Poluektov, 2013}

\begin{document}

\maketitle

\begin{abstract}
We study a superfluid Bose system with single-particle and pair
condensates on the basis of a half-phenome\-nological theory of a Bose liquid not involving the weakness
of interparticle interaction. The coupled equations describing the equilibrium state of
such system are derived from the variational principle for entropy. These equations
are analyzed at zero temperature both analytically and numerically. It is shown
that the fraction of particles in the single-particle and pair condensates essentially depends on the total density of the system. At densities attainable in condensates of alkali-metal atoms, almost all particles are in the single-particle condensate. The pair condensate fraction grows with an increasing total density and becomes dominant. It is shown that at density of liquid helium, the single-particle condensate fraction is less than 10\%,  which agrees with experimental data on inelastic neutron scattering, Monte Carlo calculations and other theoretical predictions. The ground state energy, pressure, and compressibility are found for the system under consideration. The spectrum of single-particle excitations is also analyzed.

\keywords superfluidity, Bose-Einstein condensation, single-particle and pair condensates, quasiparticles, excitation spectrum
\pacs 67.10.Fj, 67.25.D, 67.85.Jk
\end{abstract}

\section{Introduction}

The first experimental observations of Bose-Einstein condensation in dilute gases of alkali-metal atoms~\cite{Anderson,Davis,Bradley} have stimulated a great interest to this remarkable phenomenon manifested also in superfluids and superconductors.
However, in spite of a significant progress~\cite{Dalfovo,Zagrebnov,Pethick,Stringari,Andersen} in the study of Bose systems with condensate, their theory is far from being completed. Among the open theoretical problems we mention the following ones: description of Bose systems with strong interaction, microscopic justification of observable excitation spectrum in a superfluid ${}^{4}$He, role of single-particle and pair condensates in the phenomenon of superfluidity, the existence of elementary excitations with activation energy along with sound excitations.

As was first showed by Bogolyubov~\cite{Bogolyubov}, the interparticle interaction essentially affects the behavior of a many-body Bose system at low temperature. In particular, in consequence of a weak interaction, the number of particles in a condensate at zero temperature is not equal to the total particle number in the system. This effect, usually referred to as depletion of a condensate, is caused by the presence of pair anomalous averages which are similar to Cooper correlations in superconductors. The pair correlated bosonic atoms form a pair condensate which, along with a single-particle condensate, specifies the superfluid density. The role of pair correlations in superfluid Bose systems has been studied by many authors~\cite{Gross1,Girardeau,Wentzel,Luban,Poluektov2,Kobe,Shevch,NepPash,Pash,Zagrebnov1}. Superfluidity has also been treated in terms of pair condensation only, in the total absence of a single-particle condensate~\cite{Valatin,Imry,Kondr}.

Note that accounting for pair correlations in a Bose system can produce a gap in the spectrum of elementary excitations~\cite{Girardeau,Wentzel,Luban,Poluektov2}. The existence of a gap essentially depends on the used approximation or truncated Hamiltonian~\cite{Zagrebnov,Andersen,Martin,Griffin}. In particular, the well-known Hartree-Fock-Bogolyubov approximation extended to bosons generates a gap in the spectrum. In virtue of the Hugenholtz-Pines theorem~\cite{Hugenholtz}, according to which the spectrum should be gapless, this fact is usually considered as a defect of the mentioned approximation. Therefore, some efforts have been done to reformulate the Hartree-Fock-Bogolyu\-bov theory in order to remove a gap in the single-particle excitation spectrum (see~\cite{Yukalov} and references therein). However, such schemes involve additional assumptions or parameters which does not alow one to treat them as a rigorous solution of the problem. Note that Pines~\cite{Pines} has pointed out that the correctness of their theorem depends on the validity of expansions into a series of perturbation theory.
Moreover, the Hugenholtz-Pines theorem cannot be applied to all models because it is valid for a specific truncated Hamiltonian used by the authors.
The subsequent analysis of Bogolyubov's $1/k^{2}$-theorem has showed that no general conclusion can be made concerning the excitation spectrum of a superfluid and long-wavelength density excitations are insensitive to $U(1)$ symmetry breaking~\cite{Wagner}. As was also stressed by Bogolyubov and Bogolyubov (jr.), the presence of correlated pairs in a Bose gas of interacting particles leads to the fact that the spectrum consists of two branches~--- the branch with activation energy (or with a gap) and a phonon mode~\cite{Bogol-Bogol}. The similar form of spectrum for a superfluid ${}^{4}$He was also proposed within a phenomenological model~\cite{Hastings}.
Perhaps, one may consider that the existence of a gap in the spectrum of single-particle excitations does not contradict the general physical principles. Moreover, the possible existence of elementary excitations with activation energy along with sound excitations requires further theoretical and experimental investigations. The existence of a gap in the single-particle excitation spectrum has been recently discussed in terms of variational wave functions~\cite{Suto} and without the involvement of a conjecture of $c$-number representation of creation and annihilation operators with zero momentum~\cite{Trigger}. Note that the $c$-number representation results in a reduction of Fock space, though one can prove that it correctly reproduces the pressure~\cite{Ginibre} and condensate density~\cite{Lieb,Suto1} in thermodynamic limit. In a recent study~\cite{Ettouhami}, it has been shown that the restoration of the state with zero momentum to the Hilbert space with subsequent exact numerical diagonalization of the total single-mode Hamiltonian yields the excitation spectrum with a finite gap.

The description of many-body systems that does not require the weakness of interparticle interaction can be formulated using the quasiparticle concept. The well-known example of such a description is the Fermi liquid theory proposed by Landau~\cite{Landau} and Silin~\cite{Silin}. This theory has been extended to the study of various superfluid states of Fermi systems by introducing pair anomalous averages~\cite{PhysA,UFN93,PhysRep}. The main advantage of the extended half-phenomenological approach is that it is valid for an arbitrary energy functional and does not restrict the intensity of interaction. It can be applied to a wide range of systems, including strongly interacting superfluid nuclear matter~\cite{AkhIsPelRekYats,AIPY,AIPY_2}. On microscopic level, such half-phenomenological description is equivalent to a self-consistent mean field theory~\cite{Poluektov1}. Subsequently, the ideas of the extended Fermi liquid approach have also been disseminated to a Bose superfluid~\cite{PhysRep,PhysPart} with a corresponding microscopic justification~\cite{Poluektov2}. Note that the self-consistent mean field model is an effective zeroth-order approximation of quantum-field perturbative theory~\cite{Poluektov3}.

In the present paper, we study a superfluid Bose system with single-particle and pair condensates without any restriction on intensity of interaction and conjecture of $c$-number representation of creation and annihilation operators. To this end, we employ the formalism~\cite{PhysRep,PhysPart} developed by us earlier. From the principle of maximum entropy, we present a brief derivation of the coupled equations describing the system at finite temperature. The obtained equations are analyzed in detail both analytically and numerically in case of zero temperature and contact interaction. It is shown that for systems of low density (dilute gases of alkali-metal atoms), the pair condensate fraction is negligibly small in comparison with the fraction of single-particle condensate. Such systems, as expected, are well described by the Gross-Pitaevskii equation~\cite{Dalfovo,Pethick,Stringari}. The role of pair condensate becomes determinative for dense systems. In particular, it is found that the single-particle condensate fraction in a superfluid ${}^{4}$He is less than 10\% at zero temperature. This result is in a good agreement with experiments~\cite{Sosnik,Kozlov,Snow,Glyde} on inelastic neutron scattering, with Monte Carlo calculations~\cite{Moroni} and with theoretical approach that involves the calculation of single-particle density matrix expressed through the structure factor~\cite{Vakarchuk1,Vakarchuk2,Vakarchuk3}. We also calculate the ground state energy, pressure, speed of sound for a system with two condensates. The single-particle excitation spectrum is analyzed.

\clearpage

\section{Formalism and basic equations}

Consider the basic principles that underlie the theory of superfluid Bose systems with single-particle and pair condensates
(for details see the reviews~\cite{PhysRep,PhysPart}). This theory is formulated in close analogy with extended Fermi-liquid approach to superfluids~\cite{PhysA,UFN93,PhysRep}. The state of a superfluid Bose system is specified by the condensate amplitudes,
\begin{equation}\label{eq:2.1}
b_{\bf p}={\rm Sp}\,\rho a_{\bf p}\,, \qquad  b^{*}_{\bf p}={\rm Sp}\,\rho a^{\dag}_{\bf p}\,,
\end{equation}
as well as by normal and anomalous single-particle density matrices,
\begin{gather}
f_{{\bf pp}'}={\rm Sp}\,\rho a^{\dag}_{{\bf p}'}a_{\bf p}\,, \qquad
g_{{\bf pp}'}={\rm Sp}\,\rho a_{{\bf p}'}a_{\bf p}\,, \qquad g^{\dag}_{{\bf pp}'}={\rm Sp}\,\rho a^{\dag}_{{\bf p}'}a^{\dag}_{\bf p}\,, \label{eq:2.2}
\end{gather}
where creation and annihilation operators $a^{\dagger}_{\bf p}$, $a_{\bf p}$ satisfy the usual
Bose commutation relations and $\rho$ is a statistical operator of the system which we define below by equation~(\ref{eq:2.3}). Note that $f_{{\bf p}{\bf p}'}=f^{*}_{{\bf p}'{\bf p}}$ and $g_{{\bf p}{\bf p}'}=g_{{\bf p}'{\bf p}}$. The condensate amplitudes~(\ref{eq:2.1}) describe a single-particle condensate, while anomalous averages~(\ref{eq:2.2}) characterize the pair correlations between particles and indicate the existence of a pair condensate.

We will approximate the statistical operator $\rho$ by operator that contains a general quadratic form of $a_{\bf p}$, $a^{\dagger}_{\bf p}$ as well as linear terms in creation and annihilation operators,
\begin{equation}
\rho=\exp(Z-F), \qquad
F=a^{\dagger}Aa+{\frac{1}{2}}\left(aBa+a^{\dagger}B^{*}a^{\dagger}\right)+a^{\dagger}C+C^{*}a. \label{eq:2.3}
\end{equation}
Here, we have omitted the repeated summation indices bearing in mind that, e.g., $a^{\dagger}Aa\equiv a^{\dagger}_{\bf{p}}A_{\bf{pp}'}a_{\bf{p}'}$ or $a^{\dagger}C\equiv a^{\dagger}_{\bf{p}}C_{\bf{p}}$ and $Z$ is found from the normalization condition, ${\rm Sp}\,\rho=1$. The quantities $A_{\bf{pp}'}$, $B_{\bf{pp}'}$, $C_{\bf{p}}$ are related to $f_{\bf{pp}'}$, $g_{\bf{pp}'}$, $b_{\bf{p}}$ by equations~(\ref{eq:2.2}), (\ref{eq:2.1}). The terms linear in creation and annihilation operators that stand in the exponent of $\rho$ can be removed by the unitary transformation of a $c$-number shift (for details see references~\cite{PhysRep,PhysPart,MethStat}):
\begin{equation}\label{eq:2.4}
Ua_{\bf p}U^{\dag}=a_{\bf p}+b_{\bf p}\,, \qquad Ua^{\dag}_{\bf p}U^{\dag}=a^{\dag}_{\bf p}+b_{\bf p}\,.
\end{equation}
Then, the statistical operator $\varrho=U\rho U^{\dag}$ will include only quadratic terms in $a^{\dag}_{\bf p}$, $a_{\bf p}$ as well as matrices $A_{{\bf p}{\bf p}'}$, $B_{{\bf p}{\bf p}'}$,
\begin{equation}\label{eq:2.5}
\varrho=\exp\left[\tilde{Z}-a^{\dag}Aa-{\frac{1}{2}}\left(aBa+a^{\dag}B^{*}a^{\dag}\right)\right], \qquad \tilde{Z}=Z+b^{*}Ab+{\frac{1}{2}}\left(bBb+b^{*}B^{*}b^{*}\right).
\end{equation}
For this statistical operator, we have ${\rm Sp}\,\varrho a_{\bf p}=0$. Therefore, according to equations~(\ref{eq:2.4}), one obtains
$$
{\rm Sp}\,\rho a_{\bf p}={\rm Sp}\,\varrho(a_{\bf p}+b_{\bf p})=b_{\bf p}\neq 0.
$$
Using the unitary transformation (\ref{eq:2.4}) and equations (\ref{eq:2.2}), it is easy to introduce the correlation functions $f^{c}_{{\bf p}{\bf p}'}$ and $g^{c}_{{\bf p}{\bf p}'}$:
\begin{gather}
f_{{\bf p}{\bf p}'}={\rm Sp}\,\rho a^{\dag}_{{\bf p}'}a_{\bf p}=b^{*}_{{\bf p}'}b_{\bf p}+f^{c}_{{\bf p}{\bf p}'}\,, \nonumber \\
g_{{\bf p}{\bf p}'}={\rm Sp}\,\rho a_{{\bf p}'}a_{\bf p}=b_{{\bf p}'}b_{\bf p}+g^{c}_{{\bf p}{\bf p}'}\,, \label{eq:2.6}
\end{gather}
where
\begin{equation}\label{eq:2.7}
f^{c}_{{\bf p}{\bf p}'}={\rm Sp}\,\varrho a^{\dag}_{{\bf p}'}a_{\bf p}\,, \qquad g^{c}_{{\bf p}{\bf p}'}={\rm Sp}\,\varrho a_{{\bf p}'}a_{\bf p}\,.
\end{equation}
From equations (\ref{eq:2.7}), which relate the matrices $A_{{\bf p}{\bf p}'}, B_{{\bf p}{\bf p}'}$ to $f^{c}_{{\bf p}{\bf p}'}, g^{c}_{{\bf p}{\bf p}'}$, it follows~\cite{MethStat} that $\varrho=\varrho(f^{c},g^{c},g^{c\dag})$. Therefore, the entropy of a Bose system $S=-{\rm Sp}\,\rho\ln\rho=-{\rm Sp}\,\varrho\ln\varrho$ is a functional of correlation functions only, $S=S(f^{c},g^{c},g^{c\dag})$. Moreover, it can be shown~\cite{PhysRep,PhysPart} that the matrices $A_{{\bf p}{\bf p}'}$ and $B_{{\bf p}{\bf p}'}$ represent the derivatives of entropy with respect to correlation functions:
\begin{equation}\label{eq:2.8}
{\partial S\over\partial f^{c}_{{\bf p}'{\bf p}}}=A_{{\bf p}{\bf p}'}\,, \qquad {\partial S\over\partial g^{c}_{{\bf p}'{\bf p}}}={1\over 2}B_{{\bf p}{\bf p}'}\,, \qquad {\partial S\over\partial g^{c\dag}_{{\bf p}'{\bf p}}}={1\over 2}B^{\dag}_{{\bf p}{\bf p}'}\,.
\end{equation}
Taking into account equations (\ref{eq:2.8}), (\ref{eq:2.7}) we can see that these matrices are expressed through the averages of creation and annihilation operators. Therefore, they are also transformed under the global phase transformations of $a_{\bf p}$, $a^{\dag}_{\bf p}$, so that statistical operator (\ref{eq:2.5}) is invariant. In addition, such choice of $\rho$ (or~$\varrho$) is based on the fact that it satisfies the principle of spatial correlation weakening and Wick's theorem applies to it~\cite{MethStat}. The introduced statistical operator defined by equation (\ref{eq:2.5}) describes a superfluid many-body system of interacting particles in the language of free particles (or quasiparticles) with a modified dispersion law. It is worth stressing that the efficiency of such description is essentially determined by the choice of coefficients $A_{{\bf p }{\bf p}'}$ and $B_{{\bf p }{\bf p}'}$. As we will see below, these coefficients are chosen to satisfy the requirement of maximum entropy for fixed values of additive integrals of motion (or conserved quantities). In the language of a self-consistent mean field theory, this requirement is equivalent to the fact that the Hamiltonian of such approximation is the closest to the exact one~\cite{Poluektov2}. Therefore, the statistical operator (\ref{eq:2.5}) and the maximum entropy principle give the most accurate description of the system within the quasiparticle approximation even if the interparticle interaction is not weak.

A compact formulation of a theory under consideration is given in terms of a two-row matrix $\hat{f}^{c}$ that combines the correlation functions and a vector $\hat{\psi}$, whose components are the condensate amplitudes,
\begin{equation}\label{eq:2.9}
\hat{f}^{c}=\left( \begin{array}{cc}
f^{c} & -g^{c} \\ g^{c\dagger} & -1-\tilde{f}^{c}
\end{array}
\right), \qquad \hat{\psi}=\left(\begin{array}{c} b \\ b^{*}\end{array}\right),
\end{equation}
where tilde denotes the transposed matrix. Thermodynamic equilibrium of the system is determined by the maximum of entropy $S=-{\rm Sp}\,\rho\ln\rho$ for fixed values of conserved quantities, such as energy, momentum, and total particle number.
Using the unitary transformation (\ref{eq:2.4}) and $u-v$ transformations, one can show~\cite{PhysPart,PhysRep} that the statistical operator (\ref{eq:2.5}) is reduced to a diagonal form  $\rho_{0}$. Then, for the entropy of the system $S=-{\rm Sp}\,\rho_{0}\ln\rho_{0}$, the following combinatorial expression is valid:
\[
S=-{\rm tr}\,\left[f_{0}\ln f_{0}-(1+f_{0})\ln(1+f_{0})\right], \qquad f_{0}={\rm Sp}\,\rho_{0}a^{\dag}a,
\]
which allows it to be expressed through the introduced two-row matrix~\cite{PhysPart,PhysRep}:
\begin{equation}
S(\hat{f}^{c})=-{\rm Re}{\rm Tr}\, \hat{f}^{c}\ln \hat{f}^{c}. \label{eq:2.10}
\end{equation}
Here, ${\rm tr}\ldots$ is a trace over the momentum variables which specify a single-particle state, while ${\rm Tr}\ldots$ is taken over the the two-row matrix as well as the momentum variables.
The introduced entropy depends on correlation functions only and does not depend on the condensate amplitudes.

The energy of a superfluid Bose system is a functional of correlation functions and condensate amplitudes, $E=E(\hat{f}^{c},\hat{\psi})$. It is obtained by averaging a microscopic Hamiltonian,
\begin{equation} \label{eq:2.11}
E(\hat{f}^{c},\hat{\psi})={\rm Sp}\rho H\left(a^{\dag}_{\bf p},a_{\bf p}\right).
\end{equation}
In fact, the normal-ordered Hamiltonian $H\left(a^{\dag}_{\bf p},a_{\bf p}\right)$ of a Bose system can take into account binary, triple, and higher-order interactions of particles. However, we will use the following second quantized Hamiltonian with binary interparticle interaction,
\begin{equation}\label{eq:2.12}
H\left(a^{\dag}_{\bf p},a_{\bf p}\right)=\sum_{\bf p}{p^{2}\over 2m}a^{\dag}_{\bf p}a_{\bf p}+{1\over 2V}\sum_{{\bf p}_{1}\dots{\bf p}_{4}}\nu({\bf p}_{1}-{\bf
p}_{3})a^{\dag}_{{\bf p}_{1}}a^{\dag}_{{\bf p}_{2}}a_{{\bf p}_{3}}a_{{\bf p}_{4}}\delta_{{\bf p}_{1}+{\bf p}_{2},{\bf p}_{3}+{\bf p}_{4}}\,.
\end{equation}
Now, we find the explicit form of the energy functional $E(\hat{f}^{c},\hat{\psi})$ for the written Hamiltonian. Due to unitary transformation (\ref{eq:2.4}), the averaging given by equation~(\ref{eq:2.11}) is reduced to
\[E\left(\hat{f}^{c},\hat{\psi}\right)={\rm Sp}\varrho H\left(a^{\dag}_{\bf p}+b^{*}_{\bf p},a_{\bf p}+b_{\bf p}\right).
\]
Since $\varrho$ has a Gaussian form that involves both normal and anomalous pairs of second quantized operators in the exponent, we can apply Wick's theorem with non-vanishing pairwise normal and anomalous averages [or contractions, see equations (\ref{eq:2.7})]. Thus, we have~\cite{ASPSVP}
\begin{eqnarray}
E(\hat{f}^{c},\hat{\psi})&=&E(\hat{\psi})+
\sum_{{\bf p}_{1}{\bf p}_{2}}{p_{1}^{2}\over 2m}f^{c}_{{\bf p}_{1}{\bf p}_{2}}\delta_{{\bf p}_{1},{\bf
p}_{2}} \nonumber \\
&&+{1\over V}\sum_{{\bf p}_{1}\dots{\bf p}_{4}}f^{c}_{{\bf p}_{1}{\bf p}_{2}}\left[\nu({\bf p}_{1}-{\bf p }_{2})+\nu({\bf p}_{2}-{\bf p}_{4})\right] b^{*}_{{\bf p}_{3}}b_{{\bf p}_{4}}\delta_{{\bf p}_{3}+{\bf p}_{2},{\bf p}_{1}+{\bf p}_{4}}  \nonumber \\
&&+{1\over 2V}\sum_{{\bf p}_{1}\dots{\bf p}_{4}}f^{c}_{{\bf p}_{1}{\bf p}_{2}}f^{c}_{{\bf p}_{3}{\bf p}_{4}}[\nu({\bf p}_{1}-{\bf p}_{2})+
\nu({\bf p}_{2}-{\bf p}_{3})]\delta_{{\bf p}_{1}+{\bf p}_{3},{\bf p}_{2}+{\bf p}_{4}} \nonumber \\
&&+{1\over 2V}\sum_{{\bf p}_{1}\dots{\bf p}_{4}}\nu({\bf p}_{2}-{\bf p}_{3})\left[g^{c}_{{\bf p}_{1}{\bf p}_{2}}b^{*}_{{\bf p}_{3}}b^{*}_{{\bf p}_{4}}+
{\rm h.c.}\right]\delta_{{\bf p}_{1}+{\bf p}_{2},{\bf p}_{3}+{\bf p}_{4}} \nonumber \\
&&+{1\over 4V}\sum_{{\bf p}_{1}\dots{\bf p}_{4}}g^{*c}_{{\bf p}_{1}{\bf p}_{2}}g^{c}_{{\bf p}_{3}{\bf p}_{4}}[\nu({\bf p}_{1}-{\bf p}_{3})+\nu({\bf p}_{1}-{\bf p}_{4})]\delta_{{\bf p}_{1}+{\bf p}_{2},{\bf p}_{3}+{\bf p}_{4}}\,, \label{eq:2.13}
\end{eqnarray}
where the first term $E(\hat{\psi})\equiv E(b_{\bf p},b_{\bf p}^{*})$ is constructed from the condensate amplitudes $b_{\bf p}$, $b^{*}_{\bf p}$ only,
\begin{equation}\label{eq:2.14}
E(\hat{\psi})=\sum_{\bf p}{p^{2}\over 2m}b^{*}_{\bf p}b_{\bf p}+{1\over 2V}\sum_{{\bf p}_{1}\dots{\bf p}_{4}}\nu({\bf p}_{1}-{\bf
p}_{3})b^{*}_{{\bf p}_{1}}b^{*}_{{\bf p}_{2}}b_{{\bf p}_{3}}b_{{\bf p}_{4}}\delta_{{\bf p}_{1}+{\bf p}_{2},{\bf p}_{3}+{\bf p}_{4}}\,.
\end{equation}
\looseness=-1It can be proved that $E(\hat{\psi})$ is related to the total energy functional $E(\hat{f}^{c},\hat{\psi})$ by the differential operator~\cite{PhysRep,PhysPart},
\[
E(\hat{f}^{c},\hat{\psi})=RE(\hat{\psi}),
\]
where
\[
R=\exp\left({\partial\over\partial b}f^{c}{\partial\over\partial b^{*}}+{1\over 2}{\partial\over\partial b}g^{c}{\partial\over\partial b}+{1\over 2}{\partial\over\partial b^{*}}g^{c*}{\partial\over\partial b^{*}}\right).
\]
In the expression for $R$, we have omitted the repeated summation indices, like in equations~(\ref{eq:2.3}), (\ref{eq:2.5}).

The total particle number, similar to entropy (\ref{eq:2.10}), can be expressed through the two-row matrix $\hat{f}^{c}$. Like in quantum mechanics, in which the physical quantities correspond to operators, we assume that a physical quantity $a$ is associated with a two-row matrix $\hat{a}$. Then, the average of a physical quantity $a$ is given by~\cite{PhysPart,PhysRep}
\[
\langle a\rangle\equiv {\rm tr}\,fa={1\over 2}\left({\rm Tr}\, \hat{f}^{c}\hat{a}-{\rm tr}\,a+\hat{\psi}^{*}\hat{\tau}_{3}\hat{a}\hat{\psi}\right), \qquad \hat{a}=\left(\begin{array}{cc} a & 0 \\ 0 & -\tilde{a}\end{array}\right),
\]
where $\hat{\tau}_{3}$ is the Pauli matrix and ${\rm tr}\, fa\equiv\sum_{\bf{p}}(fa)_{\bf{pp}}$. Since $\hat{\tau}_{3}$ is the generator of unitary gauge transformations, it should be interpreted as the particle number operator. Therefore, according to the above formula, we have
\[
N\left(\hat{f}^{c},\hat{\psi}\right)= {\rm tr}\,f={1\over 2}\left({\rm Tr}\, \hat{f}^{c}\hat{\tau}_{3}-{\rm tr}\,1+\hat{\psi}^{*}\hat{\psi}\right), \qquad \hat{\tau}_{3}=\left(\begin{array}{cc} 1 & 0 \\ 0 & - 1\end{array}\right),
\]
where $1\equiv \delta_{\bf{pp}'}$. The calculation of the traces gives the following expression for the total particle number:
\begin{equation} \label{eq:2.15}
N\left(\hat{f}^{c},\hat{\psi}\right)=\sum_{\bf p}\left(f^{c}_{\bf p}+b^{*}_{\bf p}b_{\bf p}\right), \qquad f^{c}_{\bf pp}\equiv f^{c}_{\bf p}\,.
\end{equation}

As we have already mentioned, equations that determine the equilibrium values of correlation functions and condensate amplitudes are obtained from the principle of maximum entropy for fixed values of the additive integrals of motion~--- energy, total particle number, and total momentum. However, below we will study the system at rest (we do not introduce the latter integral of motion). Then, the problem of conditional maximization of the entropy can be reduced to the problem of unconditional minimization of the following non-equilibrium thermodynamic potential:
\begin{equation}
\Omega\left(\hat{f}^{c},\hat{\psi}\right)=-S\left(\hat{f}^{c}\right)+\beta\left[E\left(\hat{f}^{c},\hat{\psi}\right)
-\mu N\left(\hat{f}^{c},\hat{\psi}\right)\right], \label{eq:2.16}
\end{equation}
where $\beta$, $\beta\mu$ are the corresponding Lagrange multipliers ($\beta=1/T$ is the reciprocal temperature and $\mu$ is the chemical potential). The solution of the formulated variational problem gives the self-consistency equations~\cite{PhysPart,PhysRep}:
\begin{gather}
\hat{f}^{c}=[\exp\beta(\hat{\varepsilon}-\hat{\mu})-1]^{-1}, \label{eq:2.17} \\
\hat{\eta}-\mu\hat{\psi}=0, \label{eq:2.17'}
\end{gather}
where
\begin{equation} \label{eq:2.18}
\hat{\varepsilon}=\left(\begin{array}{cc}\varepsilon & \Delta \\ -\Delta^{*} & -\tilde{\varepsilon}\end{array}\right), \qquad \hat{\mu}=\left(\begin{array}{cc} \mu & 0 \\ 0 & -\mu \end{array}\right), \qquad \hat{\eta}=\left(\begin{array}{c} \eta \\ \eta^{*} \end{array}\right)
\end{equation}
and
\begin{gather}
\varepsilon_{{\bf pp}'}={\partial E\left(\hat{f}^{c},\hat{\psi}\right)\over\partial f^{c}_{{\bf p}'{\bf p}}}\,, \qquad \Delta_{{\bf pp}'}=2{\partial E\left(\hat{f}^{c},\hat{\psi}\right)\over\partial g^{c*}_{{\bf p}'{\bf p}}}\,, \qquad \eta_{{\bf p}}={\partial E\left(\hat{f}^{c},\hat{\psi}\right)\over\partial b^{*}_{\bf p}}\,.  \label{eq:2.19}
\end{gather}
Equation (\ref{eq:2.17}) has a natural form~--- namely, it reflects the fact that the matrix $\hat{f}^{c}$, which is constructed from the correlation functions, has a structure of the Bose distribution function if $\hat{\varepsilon}$ is interpreted as the operator of quasiparticle energy. Note that $\hat{\varepsilon}$ and $\hat{f}^{c}$ are Hermitian matrices in indefinite metrics introduced in references~\cite{PhysRep,PhysPart}. Equation (\ref{eq:2.17'}) has a structure similar to the stationary Gross-Pitaevskii equation without external potential. However, the principal difference is that the energy of the system depends now not only on the condensate amplitudes but also on correlation functions. The coupled equations (\ref{eq:2.17}), (\ref{eq:2.17'}) describe an inhomogeneous equilibrium state of a superfluid Bose system with single-particle and pair condensates if the energy functional $E\left(\hat{f}^{c},\hat{\psi}\right)$ is known.

In conclusion of this section we would like to note that a similar approach has been developed to extend the theory for a normal Fermi liquid to superfluid states~\cite{PhysA,UFN93,PhysRep}. In this
case $\langle a_{\bf p}\rangle=\langle a^{\dagger}_{\bf{p}}\rangle=0$ and the system is described by one equation only, which has a structure of the Fermi distribution function [similar to equation~(\ref{eq:2.17})].

\section{Spatially homogeneous state}

In the case of a homogeneous system, the self-consistency equations (\ref{eq:2.17})--(\ref{eq:2.19}) have a more simple form. In particular, the correlation functions and condensate amplitudes meet the following relations:
\begin{equation} \label{eq:3.1}
f^{c}_{{\bf pp}'}=f^{c}_{\bf p}\delta_{{\bf p},{\bf p}'}\,, \qquad g^{c}_{{\bf pp}'}=g^{c}_{\bf p}\delta_{{\bf p},-{\bf p}'}\,, \qquad b_{{\bf p}}=b_{0}\delta_{{\bf p},0}\,,
\end{equation}
where $f^{c}_{\bf p}=f^{c*}_{\bf p}$ and $g_{\bf p}^{c}=g^{c}_{-{\bf p}}$. These relations show that only particles with zero momentum and pairs of particles with total zero momentum are in a condensate. Then, in accordance with equations~(\ref{eq:2.19}), we have
\begin{equation} \label{eq:3.2}
\varepsilon_{{\bf pp}'}=\varepsilon_{\bf p}\delta_{{\bf p},{\bf p}'}\,, \qquad \Delta_{{\bf pp}'}=\Delta_{\bf p}\delta_{{\bf p},-{\bf p}'}\,,
\end{equation}
where
\begin{equation} \label{eq:3.2'}
\varepsilon_{\bf p}={\partial E\left(\hat{f}^{c},\hat{\psi}\right)\over\partial f^{c}_{\bf p}}\,, \qquad  \Delta_{\bf p}={\partial E\left(\hat{f}^{c},\hat{\psi}\right)\over\partial g^{c*}_{\bf p}}\,.
\end{equation}

Let us transform equation (\ref{eq:2.17}) taking into account the above conditions for spatial homogeneity. To this end, we expand the $2\times2$ matrix $\beta\hat{\xi}=\beta\left(\hat{\varepsilon}-\hat{\mu}\right)$, entering this equation, in terms of the traceless Pauli matrices $\hat{\tau}_{i}$ and the identity matrix $\hat{\tau}_{0}$:
\begin{equation} \label{eq:3.3}
\beta\hat{\xi}=\beta\left(\begin{array}{cc}\xi & \Delta \\ -\Delta^{*} & -\xi\end{array}\right)=c_{i}\hat{\tau}_{i}+c_{0}\hat{\tau}_{0}\,,
\end{equation}
where $\xi=\varepsilon-\mu$. We have employed the fact that for a spatially homogeneous state the first relation from (\ref{eq:3.2}) yields $\tilde{\xi}=\xi$. Taking the trace of both sides of equation (\ref{eq:3.3}) we have $c_{0}=0$ and, consequently,
\[
(c_{i}\tau_{i})^{2}=\beta^{2}E_{\bf p}^{2}\,, \qquad E_{\bf p}^{2}=\xi_{\bf p}^{2}-\Delta_{\bf p}\Delta^{*}_{\bf p}\,.
\]
Let $n(x)=(e^{x}-1)^{-1}$ be the Bose distribution function. Then, equation (\ref{eq:2.17}) can be written in the form $\hat{f}^{c}=n\left(\beta\hat{\xi}\right)$. Bearing in mind equation (\ref{eq:3.3}), we can find the following representation for this equation:
\[
\hat{f}^{c}={1\over 2}\left[n\left(\beta E_{\bf p}\right)+{n\left(-\beta E_{\bf p}\right)}\right]+{1\over 2E_{\bf p }}\left[n\left(\beta E_{\bf p}\right)-{n\left(-\beta E_{\bf p}\right)}\right]\hat{\xi}.
\]
Finally, the comparison of the matrix elements of both sides of this equation yields,
\begin{equation}\label{eq:3.4}
f^{c}_{\bf p}=-{1\over 2}+{\xi_{\bf p}\over 2E_{\bf p}}\left[1+2n\left(\beta E_{\bf p}\right)\right], \qquad
g^{c}_{\bf p}=-{\Delta_{\bf p}\over 2E_{\bf p}}\left[1+2n\left(\beta E_{\bf p}\right)\right],
\end{equation}
where we have used equations (\ref{eq:3.1}), (\ref{eq:3.2}) and the evident property $n(-x)=-1-n(x)$. Here,
\begin{equation} \label{eq:3.4'}
\xi_{\bf p}=\varepsilon_{\bf p}-\mu\,, \qquad E_{\bf p}=\sqrt{\xi^{2}_{\bf p}-|\Delta_{\bf p}|^{2}}
\end{equation}
and $\varepsilon_{\bf{p}}$, $\Delta_{\bf p}$ are determined by the energy functional according to equations (\ref{eq:3.2'}).

In the case of a homogeneous system, equation (\ref{eq:2.17'}) is reduced to
\begin{equation}\label{eq:3.5}
{\partial E\left(\hat{f}^{c},\hat{\psi}\right)\over \partial b^{*}_{0}}-\mu b_{0}=0.
\end{equation}
Equations (\ref{eq:3.4})--(\ref{eq:3.5}) provide a complete description of a homogeneous Bose system with single-particle and pair condensates if the energy functional $E\left(\hat{f}^{c},\hat{\psi}\right)$ is given. The quantities $b_{0}$ and $\Delta_{\bf p}$ should be interpreted as the order parameters associated with single-particle and pair condensates, respectively, and $E_{\bf p}$ as the quasiparticle energy.

Let us now obtain the system of coupled equations using an explicit form of the energy functional given by equations (\ref{eq:2.13}), (\ref{eq:2.14}). In this case, equation (\ref{eq:3.5}) for the condensate amplitudes takes the form
\begin{equation} \label{eq:3.6}
b_{0}\left[n_{0}\nu(0)-\mu\right]+{b_{0}\over V}\sum_{\bf p}f^{c}_{\bf p}\left[\nu(0)+\nu\left({\bf p}\right)\right]+{b^{*}_{0}\over V}\sum_{\bf p}g^{c}_{\bf p}\nu({\bf p})=0,
\end{equation}
where we have used equations (\ref{eq:3.1}) and introduced the single-particle condensate density  $n_{0}=b_{0}^{*}b_{0}/V$. Next, using equations~(\ref{eq:3.2'}), we can find the quantities $\varepsilon_{\bf p}$ and $\Delta_{\bf p}$ through $f^{c}_{\bf p}=f^{c*}_{\bf p}$ and $g^{c}_{\bf p}=g^{c}_{-{\bf p}}$. The subsequent substitution of equilibrium correlation functions (\ref{eq:3.4}) gives equations for $\xi_{\bf p}=\varepsilon_{\bf p}-\mu$ and $\Delta_{\bf p}$. We should also eliminate the correlation functions in equation (\ref{eq:3.6}). After some algebraic manipulations, we come to the desired system of coupled equations for determining $\xi_{\bf p}$, $\Delta_{\bf p}$, $b_{0}$:
\begin{eqnarray}
&&\xi_{\bf p}={p^{2}\over 2m}-\mu+n_{0}\left[\nu(0)+\nu({\bf p})\right]-{1\over 2V}\sum_{{\bf p}'}\left[\nu(0)+\nu\left({\bf p}'-{\bf p}\right)\right]\left\{1-{\xi_{{\bf p}'}\over E_{{\bf p}'}}\left[1+2n\left(\beta E_{{\bf p}'}\right)\right]\right\}\,, \nonumber \\
&&\Delta_{\bf p}={1\over V}\nu({\bf p})b_{0}^{2}-{1\over 2V}\sum_{{\bf p}'}\nu\left({\bf p }+{\bf p}'\right){\Delta_{{\bf p}'}\over E_{{\bf p}'}}\left[1+2n\left(\beta E_{{\bf p}'}\right)\right]\,, \nonumber \\
&&b_{0}\left(2n_{0}\nu(0)-\xi_{0}-{b_{0}^{*}\over b_{0}}\Delta_{0}\right)=0\,. \label{eq:3.7}
\end{eqnarray}
Note that these equations, as well as equation (\ref{eq:3.6}), are invariant with respect to the following transformations: $b_{0}\to b_{0}'=b_{0}e^{\ri\varphi}$, $\xi_{\bf p}\to\xi'_{\bf p}=\xi_{\bf p}$, $\Delta_{\bf p}\to\Delta'_{\bf p}=\Delta_{\bf p}e^{2\ri\varphi}$. The same system of the coupled equations has been also derived within other approaches~\cite{Poluektov2,Tolmachev}. It has three types of solutions. The first one, with $n_{0}=0$ and $\Delta_{\bf p}=0$, describes the state with no broken symmetry. In this normal state there is neither single-particle nor pair condensate. The second type of solution, with $n_{0}=0$ and $\Delta_{\bf p}\neq 0$, breaks $U(1)$ symmetry and corresponds to the state with a pair condensate, which is similar to a condensate of Cooper pairs in the theory of superconductivity. This condensation has been studied by a number of authors~\cite{Valatin,Imry,Kondr,Thouless}. Finally, the third kind of solution, with $n_{0}\neq 0$ and $\Delta_{\bf p}\neq 0$, characterizes the state with broken $U(1)$ symmetry containing both single-particle and pair condensates. Note that the derived equations have no solution of the type $\Delta_{\bf p}=0$, $n_{0}\neq 0$.

The chemical potential is related to the total particle number. From equations (\ref{eq:2.15}), (\ref{eq:3.4}), we find that the total particle density can be written in the form
\begin{equation} \label{eq:3.8}
n=n_{0}+n_{\rm pair}+n_{\rm quas}\,,
\end{equation}
where
\begin{eqnarray}
n_{\rm pair}&=&{1\over 2V}\sum_{\bf p}\left({\xi_{\bf p}\over E_{\bf p}}-1\right)\left[1+2n\left(\beta E_{\bf p}\right)\right], \label{eq:3.8'} \\
n_{\rm quas}&=&{1\over V}\sum_{\bf p}n\left(\beta E_{\bf p}\right) \label{eq:3.8''}.
\end{eqnarray}
Here, we have introduced the particle density in a pair condensate (pair condensate density) $n_{\rm pair}$ and quasiparticle density $n_{\rm quas}$. In the normal phase, i.e., in the absence of both condensates, the particle number coincides with the quasiparticle number, as in the normal Fermi liquid. In a superfluid phase, the particle number is always greater than quasiparticle number.

\section{Zero temperature and contact interaction}

The coupled equations (\ref{eq:3.7}) can be further simplified in the case of zero temperature and contact interaction, $\nu({\bf p})=\nu={\rm const}$. Indeed, in this case, there are no quasiparticle excitations, $n\left(\beta E_{\bf p}\right)=0$, and all particles are in the single-particle or pair condensates. Moreover, due to contact interaction, $\Delta_{\bf p}$ does not depend on momentum, $\Delta_{\bf
p}\equiv\Delta$, and the structure of equations allows us to consider $b_{0}$ and $\Delta$ as real and positive quantities. After simple transformations, equations (\ref{eq:3.7}) are reduced to
\begin{eqnarray}
&&{\nu\over V}\sum_{\bf p}\left({\xi_{\bf p}\over E_{\bf p}}-1\right)-\mu+\Delta  =  0, \label{eq:4.1} \\
&&\Delta\left(1+{\nu\over 2V}\sum_{\bf p}{1\over E_{\bf
p}}\right)-\nu n_{0}  =  0, \label{eq:4.2}
\end{eqnarray}
where
\begin{equation} \label{eq:4.3}
\xi_{\bf p}={p^{2}\over 2m}+\alpha, \qquad \alpha=2\nu n_{0}-\Delta.
\end{equation}
The spectrum of elementary excitations [see equations (\ref{eq:3.4'})] takes the form
\begin{equation}\label{eq:4.4}
E_{\bf p}=\sqrt{\left({p^{2}\over 2m}+\alpha\right)^{2}-\Delta^{2}}.
\end{equation}
As we see, it has a gap $\delta\equiv E_{{\bf p}=0}$ given by
\begin{equation}\label{eq:4.5}
\delta=\sqrt{\alpha^{2}-\Delta^{2}}=\sqrt{4\nu n_{0}(\nu n_{0}-\Delta)}.
\end{equation}
We will address the issue of single-particle excitation spectrum and its behavior in the region of small momenta herein below.

Equations (\ref{eq:4.1}), (\ref{eq:4.2}) allow one to find $n_{0}$ and $\Delta$ as functions of chemical potential [or taking into account equations (\ref{eq:3.8}), (\ref{eq:3.8'}) as functions of total density]. As we have already mentioned, besides the solution corresponding to the normal state, these equations describe the state with a pair condensate only as well as the state with both condensates. Equation (\ref{eq:4.2}) shows that the solution corresponding to the state with a pair condensate ($n_{0}=0$, $\Delta\neq 0$) exists only in the case of attractive interaction, $\nu<0$. However, this state was found to be thermodynamically unstable~\cite{Thouless}. Therefore, we will study the solution of equations (\ref{eq:4.1}), (\ref{eq:4.2}) that describes the state with both condensates ($n_{0}\neq 0$, $\Delta\neq 0$). Moreover, the interaction is assumed to be repulsive, $\nu>0$.

The total particle density at $T=0$, according to equations (\ref{eq:3.8}), (\ref{eq:3.8'}), becomes
\begin{equation} \label{eq:4.6}
n=n_{0}+n_{\rm pair}=n_{0}+{1\over 2V}\sum_{\bf p}\left({\xi_{\bf p}\over E_{\bf p}}-1\right).
\end{equation}
The comparison of this formula with equation (\ref{eq:4.1}) yields the expression for the chemical potential,
\begin{equation} \label{eq:4.7}
\mu=\Delta-2\nu (n_{0}-n).
\end{equation}
From equation (\ref{eq:3.4}) we can also find the normal and anomalous correlation functions at $T=0$:
\begin{equation}\label{eq:4.8}
f_{\bf p}^{c}={{\xi_{\bf p}-E_{\bf p}}\over 2 E_{\bf p}}\,, \qquad g_{\bf p}^{c}=-{\Delta\over 2E_{\bf p}}\,.
\end{equation}
The ground state energy $E^{(0)}$ is obtained by setting $T=0$ in equations (\ref{eq:2.13}), (\ref{eq:3.4}). In addition, we should take into account the conditions for spatial homogeneity (\ref{eq:3.1}) and equations (\ref{eq:4.6}), (\ref{eq:4.8}). The ground state energy density $\mathcal{E}^{(0)}=E^{(0)}/V$ is found to be
\begin{equation}\label{eq:4.9}
\mathcal{E}^{(0)}=\nu(n-n_{0})^{2}+n\Delta-{\Delta^{2}\over 2\nu}+{1\over 2V}\sum_{\bf p}\left(E_{\bf p}-\xi_{\bf p}\right)\,.
\end{equation}
When deriving this result we have eliminated the terms $\sum_{\bf p}\left(1/E_{\bf p}\right)$ and $p^{2}/2m$ using equations (\ref{eq:4.2}), (\ref{eq:4.3}), respectively.

Let us obtain an explicit expression for pressure $P=-\Omega/\beta V$. Here, $\Omega$ is the equilibrium thermodynamic potential which is found by substituting the equilibrium values of the correlation functions and condensate amplitudes into equation (\ref{eq:2.16}). Using the fact that $S=0$ at $T=0$, one obtains $P=\mu n-\mathcal{E}^{(0)}$ and, consequently,
\begin{equation}\label{eq:4.10}
P=\nu\left(n^{2}-n_{0}^{2}\right)+{\Delta^{2}\over 2\nu}+{1\over 2V}\sum_{\bf p}\left(\xi_{\bf p}-E_{\bf p}\right).
\end{equation}
Here, as well as in equation (\ref{eq:4.9}), the quantity $\xi_{\bf p}$ is given by equation (\ref{eq:4.3}). From equation (\ref{eq:4.10}) we can conclude that the pressure is positive when $\nu>0$ that is one of the stability conditions.

We now return to the coupled equations (\ref{eq:4.1}), (\ref{eq:4.2}). First, let us eliminate the chemical potential in equation (\ref{eq:4.1}) using equation (\ref{eq:4.7}). Then, we replace the summation by integration over the variable $x=p^{2}/2m$ in both equations. The integral that appears in equation (\ref{eq:4.2}) diverges at the upper limit. The physical reason for this divergence is that we have taken a delta-like contact potential with zero radius of interparticle interaction. In order to remove the divergence, we introduce a typical length scale $r_{0}$ for the range of interaction. In subsequent numerical computations, we will assume that $r_{0}$ coincides with the value of a repulsive core of the interaction potential, which is typically equal to a few angstroms. When integrating over the variable $x$ having the dimension of energy, we will cut off the divergence by $x_{0}={(\hbar k_{0})^{2}/2m}={\hbar^{2}/2mr_{0}^{2}}$. In addition, for numerical analysis of these equations it is convenient to introduce the following dimensionless quantities: $\tilde{x}=x/x_{0}$, $\tilde{\Delta}=\Delta/x_{0}$, $\tilde{\alpha}=\alpha/x_{0}$, $\tilde{n}_{0}=n_{0}r_{0}^{3}$, $\tilde{n}=nr_{0}^{3}$ and to consider $\tilde{\alpha}=4\textsl{g}\tilde{n}_{0}-\tilde{\Delta}$ [see equations (\ref{eq:4.3})] as the sought quantity, instead of $\tilde{n}_{0}$. Consequently, equations (\ref{eq:4.1}), (\ref{eq:4.2}) in dimensionless form are written as
\begin{eqnarray}
&&\tilde{n}={{\tilde{\alpha}+\tilde{\Delta}}\over 4 \textsl{g}}+{1\over 8 \pi^{2}}\int_{0}^{\infty}\rd\tilde{x}\,\sqrt{\tilde{x}}\left[{{\tilde{x}+\tilde{\alpha}}\over\sqrt{(\tilde{x}
+\tilde{\alpha})^{2}-\tilde{\Delta}^2}}-1\right], \label{eq:4.11} \\
&&{\tilde{\Delta}\over\tilde{\alpha}}\left[1+{\textsl{g}\over 2\pi^{2}}\int_{0}^{1}\rd\tilde{x}\,{\sqrt{\tilde{x}}\over\sqrt{(\tilde{x}+\tilde{\alpha})^{2}-
\tilde{\Delta}^{2}}}\right]=1, \label{eq:4.12}
\end{eqnarray}
where
\begin{equation}\label{eq:4.13}
\textsl{g}={\nu m\sqrt{2x_{0}m}\over\hbar^{3}}={\nu m\over\hbar^{2}r_{0}}
\end{equation}
is a dimensionless coupling constant and $\nu$ is related to $s$-wave scattering length $a$ by the well-known expression $\nu=4\pi\hbar^{2}a/m$. We see that the dimensionless coupling constant is determined by the ratio of the scattering length to the radius of the interaction potential,
\begin{equation}\label{eq:4.14}
\textsl{g}=4\pi{a\over r_{0}}\,.
\end{equation}
Equation (\ref{eq:4.11}) reflects the fact that the total particle density is the sum of the particle densities in the single-particle and pair condensates (the first and second terms, respectively). The coupled equations (\ref{eq:4.11}), (\ref{eq:4.12}) are characterized by two free dimensionless parameters~--- the total particle density $\tilde{n}$ and coupling constant $\textsl{g}$. Therefore, having specified these quantities, we can find $\tilde{\alpha}$ and $\tilde{\Delta}$ and thereby determine the fractions of single-particle and pair condensates as well as other characteristics of the system.

We now present the dimensionless expressions for the quasiparticle energy, pressure, and density of the ground state energy. According to equation (\ref{eq:4.4}), the single-particle excitation spectrum in terms of dimensionless quantities has the form
\begin{equation}\label{eq:4.15}
\tilde{E}_{\tilde{\bf p}}=\sqrt{\left(\tilde{p}+\tilde{\alpha}\right)^{2}-\tilde{\Delta}^{2}}\,,
\end{equation}
where $\tilde{E}_{\bf p}=E_{\bf p}/x_{0}$ and $\tilde{p}=pr_{0}/\hbar$. In order to find an appropriate expression for pressure, let us replace the summation by integration over the variable $x=p^{2}/2m$ in equation (\ref{eq:4.10}). The divergent integral, as above, is cut off by $x_{0}$. Moreover, along with the introduced dimensionless quantities, we define the dimensionless value of pressure $\tilde{P}=P/P_{0}$  where $P_{0}=\hbar^{2}/mr_{0}^{5}$. Hence,
\begin{equation}\label{eq:4.16}
\tilde{P}=\textsl{g}\left(\tilde{n}^{2}-\tilde{n}^{2}_{0}\right)+{\tilde{\Delta}^{2}\over 8\textsl{g}}+{1\over
16\pi^{2}}J_{0}\left(\tilde{\alpha},\tilde{\Delta}\right),
\end{equation}
where
\[
J_{0}\left(\tilde{\alpha},\tilde{\Delta}\right)=\int_{0}^{1}\rd\tilde{x}\sqrt{\tilde{x}}\left[\tilde{x}+
\tilde{\alpha}-\sqrt{\left(\tilde{x}+\tilde{\alpha}\right)^{2}-\tilde{\Delta}^{2}}\right].
\]
In a similar manner, one obtains the ground state energy density in a dimensionless form,
\begin{equation}\label{eq:4.17}
\tilde{\mathcal{E}}^{(0)}={1\over 2}\tilde{n}\tilde{\Delta}-{\tilde{\Delta}^{2}\over 8\textsl{g}}+\textsl{g}\left(\tilde{n}-\tilde{n}_{0}\right)^{2}-{1\over
16\pi^{2}}J_{0}\left(\tilde{\alpha},\tilde{\Delta}\right),
\end{equation}
where $\tilde{\mathcal{E}}^{(0)}=\mathcal{E}^{(0)}/P_{0}$.

In conclusion of this section, we estimate the value of $P_{0}$ for a superfluid ${}^{4}{\rm He}$.
The interaction potential of this system has a strong short-range repulsion whose radius (core) is $r_{0}=2.55$~{\AA}. The atomic mass of helium is $m=6.65\cdot 10^{-24}$~g. For these values, we find $P_{0}\approx 15.3$~atm which is somewhat less than the value of crystallization pressure $\approx 25$~atm.

\section{Weak interaction}

Here, we study the asymptotic solution of equations (\ref{eq:4.11}), (\ref{eq:4.12}) in the case of a weak interaction. These equations contain two parameters $\textsl{g}/\pi^{2}$ and $\tilde{n}\textsl{g}$, whose values are assumed to be small,
\[
{\textsl{g}\over\pi^{2}}\ll 1, \qquad \tilde{n}\textsl{g}\ll 1.
\]
As one can see from equations (\ref{eq:4.11}), (\ref{eq:4.12}), $\tilde{\alpha}=\tilde{\Delta}=0$ when $\textsl{g}=0$ and, consequently, $\tilde{\alpha}\approx\tilde{\Delta}\ll 1$ at small $\textsl{g}$. Therefore, from equation (\ref{eq:4.12}), one obtains
\[
\tilde{\Delta}=\tilde{\alpha}\left[1-{\textsl{g}\over\pi^{2}}\left(\sqrt{1+2\tilde{\alpha}}- \sqrt{2\tilde{\alpha}}\right)\right],
\]
with an accuracy of the terms of the order of $\tilde{\alpha}^{3/2}$:
\begin{equation}\label{eq:5.2}
\tilde{\Delta}=\tilde{\alpha}\left[1-{\textsl{g}\over\pi^{2}}\left(1-\sqrt{2\tilde{\alpha}}\right)\right].
\end{equation}
Next, since $\tilde{\alpha}$ and $\tilde{\Delta}$ are small, we neglect the terms $\tilde{\alpha}^{2}$ and $\tilde{\Delta}^{2}$ in equation (\ref{eq:4.11}). Then, after its integration and subsequent substitution of equation (\ref{eq:5.2}), we come to the following equation for $\tilde{\alpha}$:
\[
4\tilde{n}\textsl{g}=2\tilde{\alpha}-{\textsl{g}\over\pi^{2}}\tilde{\alpha}+{4\sqrt{2}\over 3\pi^{2}}\textsl{g}\tilde{\alpha}^{3/2}.
\]
Its approximate solution at small $\tilde{\alpha}$ is
\begin{equation}\label{eq:5.3}
\tilde{\alpha}=2\tilde{n}\textsl{g}+{\textsl{g}\over\pi^{2}}\left[\tilde{n}\textsl{g}-{8\over 3}\left(\tilde{n}\textsl{g}\right)^{3/2}\right].
\end{equation}
After the substitution of equation (\ref{eq:5.3}) into (\ref{eq:5.2}) we can obtain $\tilde{\Delta}$ with the same level of accuracy:
\begin{equation}\label{eq:5.4}
\tilde{\Delta}=2\tilde{n}\textsl{g}-{\textsl{g}\over\pi^{2}}\left[\tilde{n}\textsl{g}-{4\over 3}\left(\tilde{n}\textsl{g}\right)^{3/2}\right].
\end{equation}
The particle density in a single-particle condensate is expressed through $\tilde{\alpha}$ and $\tilde{\Delta}$ as follows: $\tilde{n}_{0}=\linebreak\left(\tilde{\alpha}+\tilde{\Delta}\right)/4\textsl{g}$ [see equation (\ref{eq:4.11})]. Therefore, the asymptotic solution given by equations (\ref{eq:5.3}), (\ref{eq:5.4}) allows us to find the particle number densities in both condensates:
\begin{equation}\label{eq:5.5}
\tilde{n}_{0}=\tilde{n}-{1\over 3\pi^{2}}\left(\tilde{n}\textsl{g}\right)^{3/2}, \qquad \tilde{n}_{\rm pair}={1\over 3\pi^{2}}\left(\tilde{n}\textsl{g}\right)^{3/2}.
\end{equation}
As we see, the terms $\sim\left(\tilde{n}\textsl{g}\right)^{3/2}$ account for the presence of a pair condensate.

The next step is to calculate the ground state energy density and pressure in the case of a weak interaction. Both quantities, according to equation (\ref{eq:4.16}), (\ref{eq:4.17}), are determined by the integral $J_{0}\left(\tilde{\alpha},\tilde{\Delta}\right)$ which at small $\tilde{\alpha}$ and $\tilde{\Delta}$ has the following asymptotic behavior:
\[
J_{0}\left(\tilde{\alpha},\tilde{\Delta}\right)\approx J_{0}\left(\tilde{\alpha}\right)\approx \tilde{\alpha}^{2}-{16\sqrt{2}\over 15}\tilde{\alpha}^{5/2}\,,
\]
or taking into account equation~(\ref{eq:5.3}),
\begin{equation}\label{eq:5.6}
J_{0}\left(\tilde{\alpha},\tilde{\Delta}\right)\approx J_{0}\left(\tilde{\alpha}\right)\approx 4\left(\tilde{n}\textsl{g}\right)^{2}-{128\over 15}\left(\tilde{n}\textsl{g}\right)^{5/2}\,.
\end{equation}
Therefore, using equations (\ref{eq:5.3})--(\ref{eq:5.5}), we come to the following dimensionless expression for pressure:
\begin{equation} \label{eq:5.7}
\tilde{P}={\tilde{n}^{2}\textsl{g}\over 2}-{1\over 4\pi^{2}}\left(\tilde{n}\textsl{g}\right)^{2}+{4\over 5\pi^{2}}\left(\tilde{n}\textsl{g}\right)^{5/2}\,.
\end{equation}
It is also easy to find the pressure in dimensional form,
\begin{equation}\label{eq:5.8}
P={\nu n^{2}\over 2}\left\{1-{1\over 2\pi^{2}}\left[{\nu m\over\hbar^{2}r_{0}}-{16\over 5}{(\nu m )^{3/2}n^{1/2}\over \hbar^{3}}\right]\right\}.
\end{equation}
Having obtained an explicit dependence of pressure on density, we can write down the speed of sound $u$:
\begin{equation}\label{eq:5.9}
u^{2}={1\over m}{\partial P\over\partial n}={\nu n\over m}\left[1-{\nu m\over 2\pi^{2} \hbar^{2}r_{0}}+{2(\nu m)^{3/2}n^{1/2}\over\pi^{2}\hbar^{3}}\right].
\end{equation}
The second term in equations (\ref{eq:5.8}), (\ref{eq:5.9}) accounts for the finite range of interaction potential, while the third term is a correction responsible for the presence of a pair condensate. The ground state energy density (\ref{eq:4.17}) is determined by the same integral $J_{0}\left(\tilde{\alpha},\tilde{\Delta}\right)$. Therefore, taking into account equations (\ref{eq:5.4})--(\ref{eq:5.6}) we have
\begin{equation}\label{eq:5.10}
\tilde{\mathcal{E}}^{(0)}={\tilde{n}^{2}\textsl{g}\over 2}-{1\over 4\pi^{2}}\left(\tilde{n}\textsl{g}\right)^{2}+{8\over 15\pi^{2}}\left(\tilde{n}\textsl{g}\right)^{5/2}\,,
\end{equation}
or in a dimensional form,
\begin{equation} \label{eq:5.11}
\mathcal{E}^{(0)}={\nu n^{2}\over 2}\left[1-{\nu m\over 2\pi^{2}\hbar^{2}r_{0}}+{16\over 15\pi^{2}}{(m\nu)^{3/2}n^{1/2}\over\hbar^{3}}\right].
\end{equation}
The ground state energy $E^{(0)}=\mathcal{E}^{(0)}V$ can be expressed in terms of the scattering length $a$,
\begin{equation}\label{eq:5.12}
E^{(0)}={2\pi a \hbar^{2}N^{2}\over mV}\left[1-{2\over\pi}{a\over r_{0}}+{128\over 15\sqrt{\pi}}\left({a^{3}N\over V}\right)^{1/2}\right].
\end{equation}
This formula almost coincides with those obtained by Lee and Yang~\cite{Lee}. The difference is in the second term which takes into account the final range of interaction potential. However, the renormalization of scattering length $a\to a\left(1-2a/\pi r_{0}\right)$ in equation (\ref{eq:5.12}) gives their original result.

Finally, we note that $\tilde{\alpha}\to\tilde{\Delta}$ in the limit $\nu\to 0$ and, consequently, a gap $\delta$ in the quasiparticle spectrum (\ref{eq:4.4}) tends to zero. In this case we come to the Bogolyubov spectrum~\cite{Bogolyubov},
\[
E_{\bf p}=\sqrt{\left({p^{2}\over 2m}\right)^{2}+{p^{2}\over m}\nu n_{0}}\;.
\]
In the same limit, the density of a single-particle condensate tends to the total particle density (the pair condensate density tends to zero).

\section{Small density and strong interaction}

Here, we study the system of small density without any restrictions on the value of interparticle interaction. To this end, it is convenient to introduce new dimensionless variables $\chi=\tilde{\Delta}/\tilde{\alpha}$ and $\eta=\tilde{\alpha}/4\tilde{n}g$ instead of $\tilde{\alpha}$ and $\tilde{\Delta}$. In virtue of their definition, these variables meet the inequalities $0<\chi,\eta<1$. Then, equations (\ref{eq:4.11}), (\ref{eq:4.12}) in terms of $\chi$ and $\eta$ read
\begin{eqnarray}
&&\eta\left(1+\chi\right)+{\textsl{g}\over\pi^{2}}\left(\tilde{n}\textsl{g}\right)^{1/2}\eta^{3/2}
J_{1}\left(\chi\right)=1, \label{eq:6.1} \\
&&{1\over\chi}-{\textsl{g}\over 2\pi^{2}}J_{2}\left(4\tilde{n}\textsl{g}\eta,4\tilde{n}\textsl{g}\eta\chi\right)=1, \label{eq:6.2}
\end{eqnarray}
where
\begin{eqnarray}
J_{1}\left(\chi\right)&=&\int_{0}^{\infty}\rd x\, \sqrt{x}\left[{{x+1}\over\sqrt{(x+1)^{2}-\chi^{2}}}-1\right], \label{eq:6.3} \\
J_{2}\left(4\tilde{n}\textsl{g}\eta,4\tilde{n}\textsl{g}\eta\chi\right)&=&\int_{0}^{1}\rd x\, {\sqrt{x}\over\sqrt{\left(x+4\tilde{n}\textsl{g}\eta\right)^{2}-\left(4\tilde{n}\textsl{g}\eta\chi\right)^{2}}}\,. \label{eq:6.4}
\end{eqnarray}
We remind the reader that the left-hand side of equation (\ref{eq:6.1}) is just the sum of the single-particle condensate fraction $n_{0}/n$ and pair condensate fraction $n_{\rm pair}/n$, respectively.

Now, assuming that the density is small we do not make any restrictions on the value of interaction,
\[
\tilde{n}\textsl{g}\ll1, \qquad {\textsl{g}\over\pi^{2}}\left(\tilde{n}\textsl{g}\right)^{1/2}\ll 1.
\]
Then, from (\ref{eq:6.1})--(\ref{eq:6.4}), we find the equations for determining $\chi$ and $\eta$ in the zeroth order in small parameters:
\[
\eta\left(1+\chi\right)-1=0, \qquad {1\over\chi}-{\textsl{g}\over\pi^{2}}-1=0,
\]
whence
\begin{equation} \label{eq:6.5}
\chi={1\over{1+\left(\textsl{g}/\pi^{2}\right)}}\,, \qquad \eta={{1+\left(\textsl{g}/\pi^{2}\right)}\over{2+\left(\textsl{g}/\pi^{2}\right)}}\,.
\end{equation}
The explicit expressions for $\chi$ and $\eta$ give $\tilde{\alpha}=4\textsl{g}\tilde{n}\eta$ and $\tilde{\Delta}=4\textsl{g}\tilde{n}\chi\eta$. Note that in this zero-order approximation, the total particle density coincides with the particle density in the single-particle condensate, $n=n_{0}$ (there is no pair condensate).

Let us now obtain the explicit form for the ground state energy and pressure. Both quantities, according to equations (\ref{eq:4.16}), (\ref{eq:4.17}), are determined by the integral $J_{0}\left(\tilde{\alpha},{\tilde{\Delta}}\right)$, which at small density is written as follows:
\[
J_{0}\left(\tilde{\alpha},\tilde{\Delta}\right)=J_{0}\left(4\tilde{n}g\eta,4\tilde{n}g\eta\chi\right) \approx(4\tilde{n}\textsl{g}\eta\chi)^2,
\]
where $\chi$ and $\eta$ are given by equations (\ref{eq:6.5}). It is easy to find that in the given approximation, the ground state energy density coincides with the pressure,
\[
\tilde{P}=\tilde{\mathcal{E}}^{(0)}={\tilde{n}^{2}\textsl{g}\over{2+g/\pi^{2}}}\,,
\]
or in a dimensional form,
\[
P=\mathcal{E}^{(0)}={\nu r_{0} (\pi\hbar n)^{2}\over {2r_{0}(\pi\hbar)^{2}+\nu m}}\,.
\]
The obtained expression for pressure allows us to calculate the speed of sound:
\[
u^{2}={1\over m}{\partial P\over\partial n}={\nu n r_{0} (\pi\hbar)^{2}\over {mr_{0}(\pi\hbar)^{2}+\nu m^{2}/2}}\,.
\]
We see that $u^{2}>0$ for repulsive interaction. This fact indicates the thermodynamic stability of the system. Next, from equation (\ref{eq:4.5}) we find the dimensionless expression for the energy gap in the single-particle excitation spectrum,
\begin{equation}
\tilde{\delta}=\sqrt{\tilde{\alpha}^{2}-\tilde{\Delta}^{2}}= 4\tilde{n}\textsl{g}\sqrt{\textsl{g}/\pi^{2}\over{2+{\textsl{g}/\pi^{2}}}}\,.
\end{equation}
In virtue of the small value of $\tilde{n}\textsl{g}$, this gap remains small even at quite strong interaction.

In conclusion of this section, let us calculate the pair condensate fraction [the second term in equation (\ref{eq:6.1})]. In the leading non-vanishing approximation over the small parameter, it has the form
\[
{n_{\rm pair}\over n}={\textsl{g}\over\pi^{2}}\left(\tilde{n}\textsl{g}\right)^{1/2}\eta^{3/2}J_{1}\left(\chi\right).
\]
Taking into account that $J_{1}\left(\chi\right)\approx (\pi/4)\chi^{2}$, where $\chi$ is given by the first formula in equations (\ref{eq:6.5}), the pair condensate fraction is found to be
\[
{\tilde{n}_{\rm pair}\over \tilde{n}}={\pi^{2}\over 4}\tilde{n}^{1/2}{\left(g/\pi^{2}\right)^{3/2}
\over{\left[1+\left(g/\pi^{2}\right)\right]^{1/2}\left[2+\left(g/\pi^{2}\right)\right]^{3/2}}}\,.
\]
It is easy to show that the obtained function increases in the range of small values of $\textsl{g}$ and decreases for ``strong'' interaction. Its maximum corresponds to the point $\textsl{g}_{*}\approx 50.9$ and does not depend on the total particle density in the approximation under consideration. This fact agrees with  numerical results shown in figure~\ref{fig:1}.

\section{Numerical results}

\subsection{Dilute gases}
\begin{figure}[!t]
\centerline{
\includegraphics[width=0.48\textwidth]{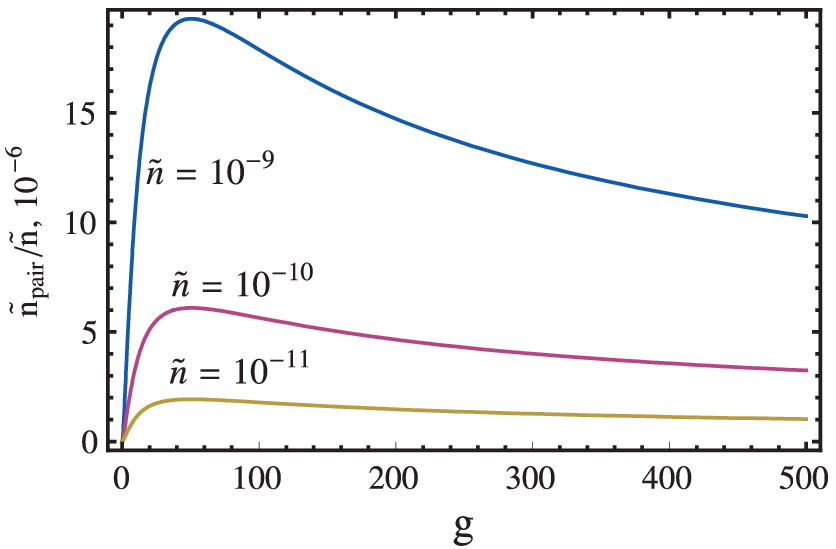}
\hfill%
\includegraphics[width=0.48\textwidth]{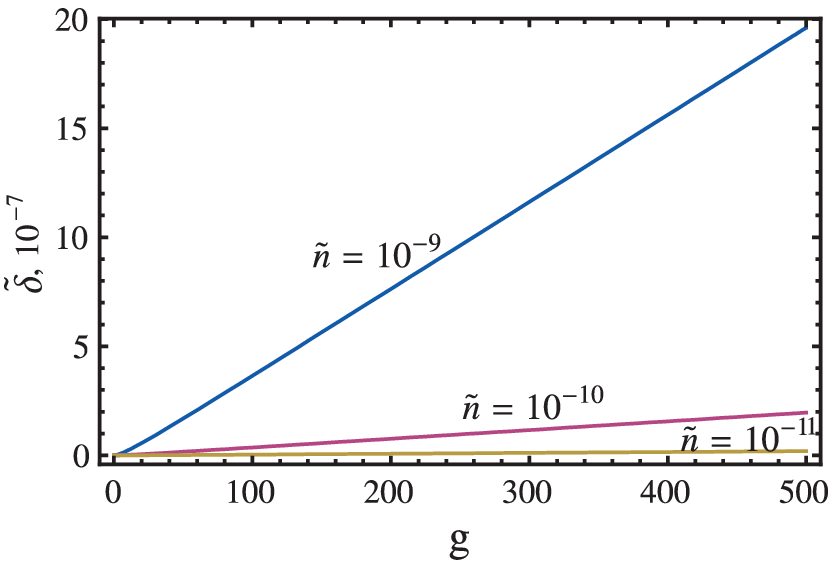}
}
\vspace{-3mm}
\parbox[t]{0.5\textwidth}{%
\caption{(Color online) Pair condensate fraction $\tilde{n}_{\rm pair}/\tilde{n}$ as a function of the coupling constant $\textsl{g}$ for various values of the total particle density $\tilde{n}$.} \label{fig:1}
}%
\parbox[t]{0.5\textwidth}{%
\caption{(Color online) Energy gap $\tilde{\delta}$ in the single-particle excitation spectrum as a function of the coupling constant $\textsl{g}$ for various values of the total particle density $\tilde{n}$.} \label{fig:2}
}%
\end{figure}

Bose-Einstein condensation in dilute ultracold gases of alkali-metal atoms was first realized experimentally in 1995~\cite{Anderson,Davis,Bradley}. The particle number density in the condensed atomic cloud is $n\sim 10^{13}\div10^{15}$~cm$^{-3}$. Then, the corresponding dimensionless density in equations (\ref{eq:6.1}), (\ref{eq:6.2}) is a small quantity, $\tilde{n}\sim 10^{-11}\div10^{-9}$, where we have taken $r_{0}\approx 3$~{\AA}. The binary atomic interaction in such systems is usually approximated by the contact interaction potential expressed through the scattering length $a$. For example, for $^{87}$Rb and $^{23}$Na, the scattering lengths are equal to $a\approx 90 a_{0}$ and $a\approx 19.1a_{0}$, respectively, where $a_{0}\approx 0.53$~{\AA} is the Bohr radius~\cite{Pethick}. From equation (\ref{eq:4.14}), we can find that the dimensionless coupling constant: $\textsl{g}\approx 200$ for $^{87}$Rb and $\textsl{g}\approx 42$ for $^{23}$Na. The numerical analysis of equations (\ref{eq:6.1}), (\ref{eq:6.2}) for a system with parameters of dilute gases of alkali-metal atoms are presented in figures~\ref{fig:1}, \ref{fig:2}.

Figure~\ref{fig:1} shows the dependencies of the pair condensate fraction $\tilde{n}_{\rm pair}/\tilde{n}$ on the coupling constant $\textsl{g}$\; for the above shown values of the total particle density $\tilde{n}$ (all quantities are dimensionless). The curves have nonmonotonous character. Moreover, their maximum is reached at $\textsl{g}_{*}\approx 50$ and practically does not depend on the total density. This result, as we have already seen in the previous section, follows from the analytical study of equations (\ref{eq:6.1}), (\ref{eq:6.2}). We can conclude that the pair condensate fraction is several orders less than the single-particle condensate fraction and, consequently, the dilute systems are described by the Gross-Pitaevskii equation with a high level of accuracy. In addition, in the limit $\textsl{g}\to 0$, the pair condensate density also tends to zero and thereby $\tilde n_{0}\to\tilde{n}$.

Figure~\ref{fig:2} presents the dimensionless energy gap $\tilde{\delta}$ in the spectrum of single-particle excitations as a function of dimensionless coupling constant $\textsl{g}$ for various values of the total particle density $\tilde{n}$. At small densities, the gap remains small even for large values of the coupling constant. The gap tends to zero in the limit $\textsl{g}\to 0$ which agrees with Bogolyubov's result~\cite{Bogolyubov}.

\subsection{Model of superfluid ${}^{4}$He}

\begin{figure}[!b]
\centerline{
\includegraphics[width=0.48\textwidth]{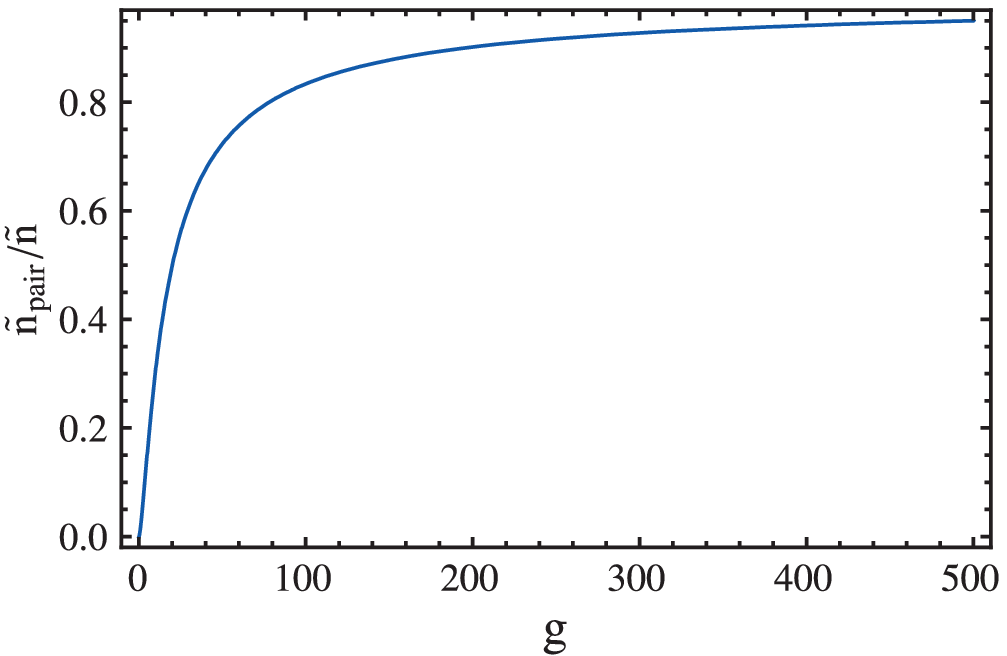}
\hfill%
\includegraphics[width=0.48\textwidth]{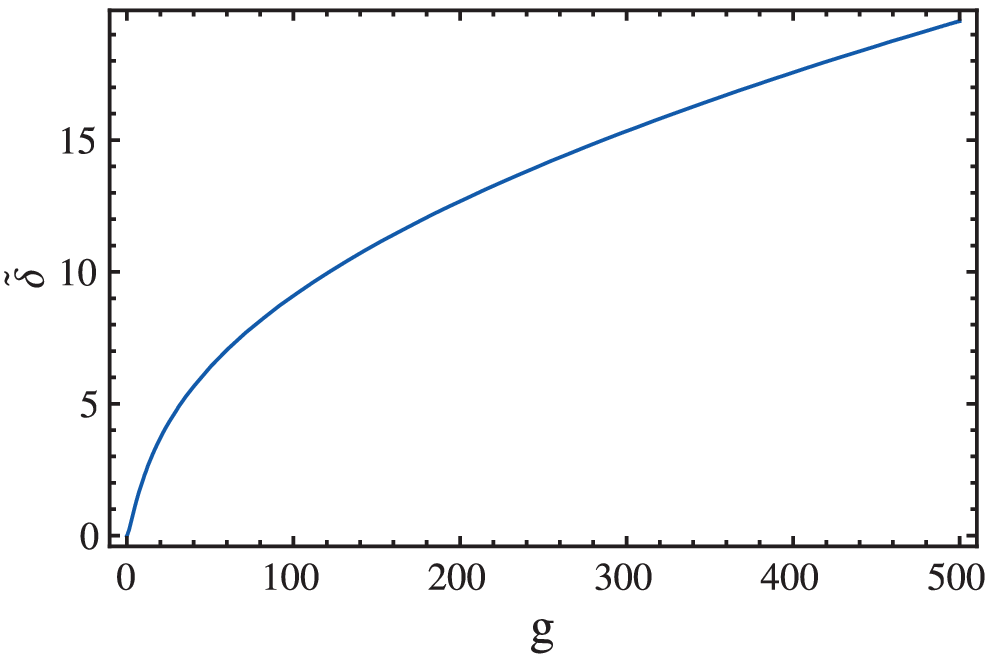}
}
\vspace{-3mm}
\parbox[t]{0.5\textwidth}{%
\caption{Pair condensate fraction $\tilde{n}_{\rm pair}/\tilde{n}$ as a function of the coupling constant $\textsl{g}$ at the density of liquid ${}^{4}$He, $\tilde{n}=0.36$.} \label{fig:3}
}%
\parbox[t]{0.5\textwidth}{%
\caption{Energy gap $\tilde{\delta}$ in the single-particle excitation spectrum as a function of the coupling constant $\textsl{g}$ at the density of liquid ${}^{4}$He, $\tilde{n}=0.36$.} \label{fig:4}
}%
\end{figure}

Superfluid ${}^{4}$He is a strongly interacting system of sufficiently high density in which Bose-Einstein condensation is also manifested~\cite{Sosnik,Kozlov,Snow,Glyde}. The density of a liquid helium at zero temperature and low pressure is about $n\sim 2.18\cdot10^{22}$~cm$^{-3}$. The atomic mass is $m=6.65\cdot 10^{-24}$~g. The interaction potential of helium has a strong short-range repulsion whose radius is $r_{0}\approx 2.55$~{\AA}, followed by a weak intermediate range attraction. Therefore, the dimensionless density $\tilde{n}=nr_{0}^3$ in equations (\ref{eq:6.1}), (\ref{eq:6.2}) is $\tilde{n}\approx 0.36$. The scattering length $a$ for $^{4}$He atoms varies from  46.1~{\AA} to 100~{\AA} depending on the interaction potential~\cite{Jamieson}. Hence, equation~(\ref{eq:4.14}) shows that the dimensionless coupling constant $\textsl{g}$ lies in the interval from 227 to 492. The numerical analysis of equations~(\ref{eq:6.1}), (\ref{eq:6.2}) for a superfluid $^{4}$He are presented in figures~\ref{fig:3}, \ref{fig:4}.

Figure~\ref{fig:3} illustrates the behavior of the pair condensate fraction depending on the coupling constant. We can see that it significantly exceeds the single-particle condensate fraction. For the above shown values of the coupling constant, the single-particle condensate fraction is just about $5\%\div10\%$. This agrees with the condensate fraction in a superfluid ${}^{4}$He measured in experiments~\cite{Sosnik,Kozlov,Snow,Glyde} and found within Monte Carlo calculations~\cite{Moroni} and theoretical predictions based on the computation of single-particle density matrix which is expressed through experimentally measured structure factor~\cite{Vakarchuk1,Vakarchuk2} (see also reference~\cite{Vakarchuk3} and references therein).

Figure \ref{fig:4} presents the dependence of the dimensionless energy gap $\tilde{\delta}$ in the single-particle excitation spectrum on $\textsl{g}$ at helium density. For example, at $\textsl{g}=227$ we have $\tilde{\delta}\approx 13.45$ or in temperature units $\delta\approx 12.5$~K. This value is somewhat greater than the value of the roton gap $\Delta_{\rm rot}\approx 8.65$~K and less than the maxon energy $\Delta_{\rm max}\approx 14$~K. The Landau gapless spectrum can be a result of superposition of excitations of various nature. Moreover, some experiments~\cite{Blag,Blag1} directed toward studying the excitations in a superfluid $^{4}$He also show the complex structure of Landau spectrum.

\section{Discussion}

Within the quasiparticle approach based on the maximum entropy principle, we have given a derivation of the coupled equations describing a superfluid Bose system with single-particle and pair condensates. These equations have been analyzed both analytically and numerically. In order to remove the divergences, we introduce a typical length scale $r_{0}$ for the range of interaction potential. For dilute systems (alkali atomic gases), the pair condensate fraction is negligibly small at zero temperature, $n_{\rm pair}/n\sim 10^{-6}$ and it grows with an increasing density $n$. For a superfluid $^{4}$He, the role of pair condensate at $T=0$ is dominant because the single-particle condensate fraction $n_{0}/n$ is less than 10\%, which  agrees with experimental data on neutron scattering in a superfluid helium~\cite{Sosnik,Kozlov,Snow,Glyde}, with Monte Carlo calculations~\cite{Moroni}, and with other theoretical predictions (see references~\cite{Vakarchuk1,Vakarchuk2,Vakarchuk3} and references therein). We have also found other characteristics of the system such as pressure, ground state energy, speed of sound, and single-particle excitation spectrum. Note that the latter exhibits a gap. We believe that this fact is sufficiently justified because the used approximation introduces quasiparticles by the general way and respects the basic principles of statistical physics.
The studied approach can also be used to derive equations of two-fluid hydrodynamics predicting the second sound wave~\cite{PhysRep,PhysPart}.
As we have already mentioned in the introduction, the analysis of Bogolyubov's $1/k^{2}$~-theorem does not provide a general conclusion concerning the excitation spectrum of a superfluid, and long-wavelength density excitations are insensitive to the gauge symmetry breaking~\cite{Wagner}. The latter fact is directly confirmed by the neutron scattering experiments in a superfluid $^{4}$He~\cite{Blag,Blag1}. The single-particle excitation spectrum exhibits a gap in the presence of a pair condensate because the pairs have the dissociation energy and, consequently, in addition to phonon branch of spectrum, there exists another branch corresponding to the excitation of pairs~\cite{Bogol-Bogol}. The same qualitative structure of the spectrum is obtained within the proposed approach. Note that the existence of a gap resembles the situation in a solid: if there is more than one atom per unit cell, both acoustic and optical branches appear. Moreover, as we know, a neutral Fermi superfluid is characterized by the single-particle excitations with a gap as well as gapless phonon mode. The separation of single-particle and collective excitations is probably less marked in a Bose system than in a Fermi system as a consequence of ``hybridization'' of these branches due to the presence of a single-particle condensate~\cite{GlydeGrif,Glyde1}. Therefore, one can expect that the Landau spectrum is a result of superposition of excitations of various nature. It is worth stressing that the experiments~\cite{Blag,Blag1} directed toward studying the excitations in a superfluid $^{4}$He also show its complex structure.

\section*{Acknowledgement}
We thank Dr. Igor Tanatarov for valuable discussions of numerical calculations.

\ukrainianpart

\title{Роль одночастинкового та парного конденсатів у бозе-системах із довільною інтенсивністю взаємодії}
\author{О.С. Пелетминський, С.В. Пелетминський, Ю.М. Полуектов}
\address{
Інститут теоретичної фізики ім. О.І. Ахієзера, ННЦ ХФТІ, вул. Академічна, 1, 61108 Харків, Україна
}

\makeukrtitle

\begin{abstract}
\tolerance=3000%
Вивчається надплинна бозе-система з одночастинковим та парним конденсатами на основі напівфеноменологічної теорії бозе-рідини, яка не припускає слабкості міжчастинкової взаємодії. Із варіаційного принципу для ентропії виведено систему рівнянь, що описують рівноважний стан такої системи. Ці рівняння проаналізовано у випадку нульової температури як аналітично, так і чисельно. Показано, що частки одночастинкового та парного конденсатів суттєво залежать від повної густини системи. При густинах, що досягаються у конденсатах атомів лужних металів, майже всі частинки знаходяться в одночастинковому конденсаті. Доведено, що при густині рідкого гелію частка одночастинкового конденсату є меншою за 10\%, що узгоджується з експериментальними даними з непружного нейтронного розсіювання, розрахунками методом Монте-Карло та іншими теоретичними передбаченнями. Знайдено вирази для енергії основного стану, тиску, стисливості системи, що вивчається. Проаналізовано також спектр одночастинкових збуджень.
\keywords надплинність, бозе-ейнштейнівська конденсація, одночастинковий та парний конденсати,
        квазічастинки, спектр збуджень

\end{abstract}


\clearpage

\begin{thebibliography}{99}

\bibitem{Anderson}
Anderson M.H., Ensher J.R., Matthews M.R., Wieman C.E., Cornell E.A., Science, 1995, \textbf{269}, 198; \\ \doi{10.1126/science.269.5221.198}.

\bibitem{Davis}
Davis K.B., Mewes M.-O., Andrews M.R., van Druten N.J., Durfee D.S., Kurn D.M.,  Ketterle W., Phys. Rev. Lett., 1995, \textbf{75}, 3969; \doi{10.1103/PhysRevLett.75.3969}.

\bibitem{Bradley}
Bradley C.C., Sackett C.A., Tollett J.J., Hulet R.G., Phys. Rev. Lett., 1995, \textbf{75}, 1687; \doi{10.1103/PhysRevLett.75.1687}.

\bibitem{Dalfovo}
Dalfovo F., Giorgini S., Pitaevskii L.P., Stringari S., Rev. Mod. Phys., 1999, \textbf{71}, 463; \doi{10.1103/RevModPhys.71.463}.

\bibitem{Zagrebnov}
Zagrebnov V.A., Bru J.B., Phys. Rep., 2001, \textbf{350}, 291; \doi{10.1016/S0370-1573(00)00132-0}.

\bibitem{Pethick}
Pethick C.J., Smith H., Bose-Einstein Condensation in Dilute Gases, Cambridge University Press, Cambridge, 2002.

\bibitem{Stringari}
Pitaevskii L., Stringari S., Bose-Einstein Condensation, Oxford Science Publications, Oxford, 2003.

\bibitem{Andersen}
Andersen J.O., Rev. Mod. Phys., 2004, \textbf{76}, 599; \doi{10.1103/RevModPhys.76.599}.

\bibitem{Bogolyubov}
Bogolyubov N.N., J. Phys. USSR, 1947, \textbf{11}, 23.

\bibitem{Gross1}
Gross E.P., Ann. Phys. (New York), 1960, \textbf{9}, 292; \doi{10.1016/0003-4916(60)90033-6}.

\bibitem{Girardeau}
Gerardeau M., Arnowitt R., Phys. Rev., 1959, \textbf{113}, 755; \doi{10.1103/PhysRev.113.755}.

\bibitem{Wentzel}
Wentzel G., Phys. Rev., 1960, \textbf{120}, 1572; \doi{10.1103/PhysRev.120.1572}.

\bibitem{Luban}
Luban M., Phys. Rev., 1962, \textbf{128}, 965; \doi{10.1103/PhysRev.128.965}.

\bibitem{Poluektov2}
Poluektov Yu.M., Low Temp. Phys., 2002, \textbf{28}, 429; \doi{10.1063/1.1491184} [Fiz. Nizk. Temp., 2002, \textbf{28}, 604 (in Russian)].

\bibitem{Kobe}
Kobe D.H., Ann. Phys., 1968, \textbf{47}, 15; \doi{10.1016/0003-4916(68)90224-8}.

\bibitem{Shevch}
Shevchenko S.I., Sov. J. Low Temp. Phys., 1985, \textbf{11}, 183 [Fiz. Nizk. Temp., 1985, \textbf{11}, 339 (in Russian)].

\bibitem{NepPash}
Nepomnyaschii Yu.A., Pashitskii E.A., Sov. Phys.--JETP, 1990, \textbf{71}, 98 [Zh. Eksp. Teor. Fiz., 1990, \textbf{98}, 178 (in Russian)].

\bibitem{Pash}
Pashitskii E.A., Low Temp. Phys., 1999, \textbf{25}, 81; \doi{10.1063/1.593709}; [Fiz. Nizk. Temp., 1999, \textbf{25}, 115 (in Russian)].

\bibitem{Zagrebnov1}
Pul\'{e} J.V., Zagrebnov V.A., Rev. Math. Phys., 2007, \textbf{19}, 157; \doi{10.1142/S0129055X07002924}.

\bibitem{Valatin}
Valatin J.G., Butler D., Nuovo Cimento, 1958, \textbf{10}, 37; \doi{10.1007/BF02859603}.

\bibitem{Imry}
Evans W.A.B., Imry Y., Nuovo Cimento B, 1969, \textbf{63}, 155; \doi{10.1007/BF02711051}.

\bibitem{Kondr}
Kondratenko P.S., Theor. Math. Phys., 1975, \textbf{22}, 196; \doi{10.1007/BF01036327} [Teor. Mat. Fiz., 1975 \textbf{22}, 278 (in Russian)].

\bibitem{Martin}
Hohenberg P.C., Martin P.C., Ann. Phys., 1965, \textbf{34}, 291; \doi{10.1016/0003-4916(65)90280-0}.

\bibitem{Griffin}
Griffin A., Phys. Rev. B, 1996, \textbf{53}, 9341; \doi{10.1103/PhysRevB.53.9341}.

\bibitem{Hugenholtz}
Hugenholtz N.M., Pines D., Phys. Rev., 1959, \textbf{116}, 489; \doi{10.1103/PhysRev.116.489}.

\bibitem{Yukalov}
Yukalov V.I., Kleinert H., Phys. Rev. A, 2006, \textbf{73}, 063612; \doi{10.1103/PhysRevA.73.063612}.

\bibitem{Pines}
Pines D., The many-body problem, Benjamin, New York, 1961.

\bibitem{Wagner}
Wagner H., Z. Phys. A, 1966, \textbf{195}, 273; \doi{10.1007/BF01325630}.

\bibitem{Bogol-Bogol}
Bogoliubov N.N., Bogoliubov N.N. (Jr.), Introduction to Quantum Statistical Mechanics, Gordon and Breach, Lausanne, 1994.

\bibitem{Hastings}
Hastings R., Halley J.W., Phys. Rev. B, 1975, \textbf{12}, 267; \doi{10.1103/PhysRevB.12.267}.

\bibitem{Suto}
S\"{u}t\H{o} A., Sz\'{e}pfalusy P., Phys. Rev. A, 2008, \textbf{77}, 023606; \doi{10.1103/PhysRevA.77.023606}.

\bibitem{Trigger}
Bobrov V.B., Trigger S.A., Yurin I.M., Phys. Lett. A, 2010, \textbf{374}, 1938; \doi{10.1016/j.physleta.2010.02.075}.

\bibitem{Ginibre}
Ginibre J., Commun. Math. Phys., 1968, \textbf{8}, 26; \doi{10.1007/BF01646422}.

\bibitem{Lieb}
Lieb E.H., Seiringer R., Yngvason J., Phys. Rev. Lett., 2005, \textbf{94}, 080401; \doi{10.1103/PhysRevLett.94.080401}.

\bibitem{Suto1}
S\"{u}t\H{o} A., Phys. Rev. Lett., 2005, \textbf{94}, 080402; \doi{10.1103/PhysRevLett.94.080402}.

\bibitem{Ettouhami}
Ettouhami A.M., Prog. Theor. Phys., 2012, \textbf{127}, 453; \doi{10.1143/PTP.127.453}.

\bibitem{Landau}
Landau L.D., Sov. Phys.--JETP, 1956, \textbf{3}, 920 [Zh. Eksp. Teor. Fiz., 1956, \textbf{30}, 1058 (in Russian)].

\bibitem{Silin}
Silin V.P., Sov. Phys.--JETP, 1958, \textbf{6}, 387 [Zh. Eksp. Teor. Fiz., 1957, \textbf{33}, 495 (in Russian)].

\bibitem{PhysA}
Krasil'nikov V.V., Peletminskij S.V., Yatsenko A.A., Physica A, 1990, \textbf{162}, 513; \doi{10.1016/0378-4371(90)90432-R}.

\bibitem{UFN93}
Akhiezer A.I., Krasil'nikov V.V., Peletminskii S.V., Yatsenko A.A., Physics-Uspekhi, 1993, \textbf{36}, 35; \\ \doi{10.1070/PU1993v036n02ABEH002127} [Usp. Fiz. Nauk, 1993, \textbf{163}, 1 (in Russian); \\ \doi{10.3367/UFNr.0163.199302a.0001}].

\bibitem{PhysRep}
Akhiezer A.I., Krasil'nikov V.V., Peletminskii S.V., Yatsenko A.A., Phys. Rep., 1994, \textbf{245}, 1; \\ \doi{10.1016/0370-1573(94)90060-4}.

\bibitem{AkhIsPelRekYats}
Akhiezer A.I., Isaev A.A., Peletminskii S.V., Rekalo A.P., Yatsenko A.A., J. Exp. Theor. Phys., 1997, \textbf{85}, 1; \doi{10.1134/1.558307} [Zh. Eksp. Teor. Fiz., 1997, \textbf{112}, 3 (in Russian)].

\bibitem{AIPY}
Akhiezer A.I., Isayev A.A., Peletminsky S.V., Yatsenko A.A., Phys. Lett. B, 1999, \textbf{451}, 430; \\ \doi{10.1016/S0370-2693(99)00241-5}.

\bibitem{AIPY_2}
Akhiezer A.I., Isayev A.A., Peletminsky S.V., Yatsenko A.A.,
Phys. Rev. C, 2001, \textbf{63}, 021304(R); \\ \doi{10.1103/PhysRevC.63.021304}.

\bibitem{Poluektov1}
Poluektov Yu.M., Ukr. J. Phys., 2005, \textbf{50}, 1303.

\bibitem{PhysPart}
Krasil'nikov V.V., Peletminskii S.V., Phys. Part. Nucl., 1993, \textbf{24}, 200 [Fiz. Elem. Chastits At. Yadra, 1993, \textbf{24}, 463 (in Russian)].

\bibitem{Poluektov3}
Poluektov Yu.M., Ukr. J. Phys., 2007, \textbf{52}, 579.

\bibitem{Sosnik}
Sosnick T.R., Snow W.M., Sokol P.E., Silver R.N., Europhys. Lett., 1989, \textbf{9}, 707; \doi{10.1209/0295-5075/9/7/016}.

\bibitem{Kozlov}
Bogoyavlenskii I.V., Karnatsevich L.V., Kozlov Sh.A., Puchkov A.V., Sov. J. Low Temp. Phys., 1990,
\textbf{16}, 77 [Fiz. Nizk. Temp., 1990, \textbf{16}, 139 (in Russian)].

\bibitem{Snow}
Snow W.M., Sokol P.E., J. Low Temp. Phys., 1995, \textbf{101}, 881; \doi{10.1007/BF00754515}.

\bibitem{Glyde}
Glyde H.R., Diallo S.O., Azuah R.T., Kirichek O., Taylor J.W., Phys. Rev. B, 2011, \textbf{83}, 100507(R); \\ \doi{10.1103/PhysRevB.83.100507}.

\bibitem{Moroni}
Moroni S., Boninsegni M., J. Low. Temp. Phys., 2004, \textbf{136}, 129; \doi{10.1023/B:JOLT.0000038518.10132.30}.

\bibitem{Vakarchuk1}
Vakarchuk I.A., Theor. Math. Phys., 1985, \textbf{65}, 1164; \doi{10.1007/BF01017941} [Teor. Mat. Fiz., 1985 \textbf{65}, 285 (in Russian)];

\bibitem{Vakarchuk2}
Vakarchuk I.A., Theor. Math. Phys., 1990, \textbf{82}, 308; \doi{10.1007/BF01029225} [Teor. Mat. Fiz., 1990 \textbf{82}, 438 (in Russian)].

\bibitem{Vakarchuk3}
Rovenchak A.A., Vakarchuk I.O., J. Phys. Stud., 2007, \textbf{11}, 404.

\bibitem{MethStat}
Akhiezer A.I., Peletminskii S.V., Methods of Statistical Physics, Pergamon Press, Oxford, 1981.

\bibitem{ASPSVP}
Peletminskii A.S., Peletminskii S.V., Low Temp. Phys., 2010, \textbf{36}, 693; \doi{10.1063/1.3490834} [Fiz. Nizk. Temp., 2010, \textbf{36}, 875 (in Russian)].

\bibitem{Tolmachev}
Tolmachev V.V., Theory of a Bose Gas, Moscow University Press, Moscow, 1969 (in Russian).

\bibitem{Thouless}
Jeon G.S., Yin L., Rhee S.W., Thouless D.J., Phys. Rev. A, 2002, \textbf{66}, 011603(R); \doi{10.1103/PhysRevA.66.011603}.

\bibitem{Lee}
Lee T.D., Yang C.N., Phys. Rev., 1957, \textbf{105}, 1119; \doi{10.1103/PhysRev.105.1119}.

\bibitem{Jamieson}
Jamieson M.J., Dalgarno A., Kimura M., Phys. Rev. A, 1995, \textbf{51}, 2626; \doi{10.1103/PhysRevA.51.2626}.

\bibitem{Blag}
Blagoveshchenskii N.M., Bogoyavlenskii I.V., Karnatsevich L.V., Kozlov Zh.A., Kolobrodov V.G., Puchkov~A.V., Skomorokhov~A.N., JETP Lett., 1993, \textbf{57}, 428 [Pis'ma Zh. Eksp. Teor. Fiz., 1993, \textbf{57}, 414 (in Russian)].

\bibitem{Blag1}
Blagoveshchenskii N.M., Bogoyavlenskii I.V., Karnatsevich L.V., Kozlov Zh.A., Kolobrodov~V.G., Priezzhev~V.B., Puchkov~A.V., Skomorokhov~A.N., Yarunin~V.S., Phys. Rev. B, 1994, \textbf{50}, 16550; \doi{10.1103/PhysRevB.50.16550}.

\bibitem{GlydeGrif}
Glyde H.R., Griffin A., Phys. Rev. Lett., 1990, \textbf{65}, 1454; \doi{10.1103/PhysRevLett.65.1454}.

\bibitem{Glyde1}
Glyde H.R., Phys. Rev. B, 1992, \textbf{45}, 7321; \doi{10.1103/PhysRevB.45.7321}.


\end{thebibliography}
\end{document}